\documentstyle[emulateapj,psfig,apjfonts]{article}
\submitted{Received 2009 June 25; Accepted: 2009 August 6}

\def\gs{\mathrel{\raise0.35ex\hbox{$\scriptstyle >$}\kern-0.6em
\lower0.40ex\hbox{{$\scriptstyle \sim$}}}}
\def\ls{\mathrel{\raise0.35ex\hbox{$\scriptstyle <$}\kern-0.6em
\lower0.40ex\hbox{{$\scriptstyle \sim$}}}}

\makeatother

\lefthead{Smail et al.}
\righthead{Extended X-ray emission around 4C\,60.07 at $z=3.79$}

\begin{document}

\title{A 100-kpc inverse Compton X-ray halo around 4C\,60.07 at $z=3.79$}

\author{
Ian Smail,\altaffilmark{1} 
B.\,D.\ Lehmer,\altaffilmark{1} R.\,J.\ Ivison,\altaffilmark{2,3} D.\,M.\ Alexander,\altaffilmark{1}  R.\,G.\ Bower,\altaffilmark{1}\\
J.\,A.\ Stevens,\altaffilmark{4}  J.\,E.\ Geach,\altaffilmark{1} C.\,A.\ Scharf,\altaffilmark{5}
 K.\,E.\,K.\ Coppin\altaffilmark{1} \& W.\,J.\,M.\ van Breugel\altaffilmark{6} 
}

\altaffiltext{1}{Institute for Computational Cosmology, Department of Physics, Durham University, South Road, Durham DH1 3LE UK}
\altaffiltext{2}{Astronomy Technology Centre, Royal Observatory, Blackford Hill, Edinburgh EH9 3HJ}
\altaffiltext{3}{Scottish University Physics Alliance, Institute for Astronomy, University of Edinburgh, Blackford Hill, Edinburgh EH9
3HJ, UK}
\altaffiltext{4}{Centre for Astrophysics Research, Science and Technology Research Centre, University of Hertfordshire, College Lane, Herts AL10 9AB, UK}
\altaffiltext{5}{Columbia Astrophysics Laboratory, Columbia University, 550 W.\ 120th Street, MC\,5247, New York, NY 10027, USA }
\altaffiltext{6}{University of California, Merced, PO Box 2039, Merced, CA 95344, USA}

\setcounter{footnote}{4}

\begin{abstract}
We analyse a 100-ks {\it Chandra} observation of the powerful radio galaxy, 4C\,60.07 at $z=3.79$.  We identify extended X-ray emission with $L_{\rm X}\sim 10^{45}$\,erg\,s$^{-1}$ across a  $\sim 90$-kpc region around the radio galaxy. The energetics of this X-ray halo and its morphological similarity to the radio emission from the galaxy suggest that it arises from inverse Compton (IC) scattering, by relativistic electrons in the radio jets, of Cosmic Microwave Background photons and potentially far-infrared photons from the dusty starbursts around this galaxy.   The X-ray emission has a similar extent and morphology to the  Ly-$\alpha$ halo around the galaxy, suggesting that it may be ionising this halo. Indeed we find that the GHz-radio and X-ray and Ly-$\alpha$ luminosities of the halo around 4C\,60.07 are identical to those of  4C\,41.17 (also at $z=3.8$) implying that these three components are linked by a single physical process.    This is only the second example of highly-extended IC emission known at $z>3$, but it underlines the potential importance of IC emission in the formation of the most massive galaxies at high redshifts.  In addition, we detect two  X-ray luminous active galactic nuclei (AGN) within $\sim 30$\,kpc of the radio galaxy.  These two companion AGN imply that the radio and starburst activity in the radio galaxy is triggered through multiple mergers of massive progenitors on a short timescale, $\ls 100$\,Myrs.  These discoveries demonstrate the wealth of information which sensitive X-ray observations can yield into the formation of massive galaxies at high redshifts.
\end{abstract}


\section{Introduction}

Recipes for feedback from AGN have been introduced into galaxy formation and evolution models to counter the catastrophic over-cooling of gas (and hence excess star formation) in massive halos (e.g.\ Benson et al.\ 2003; Bower et al.\ 2006; Sijacki et al.\ 2007).  Such processes can also potentially explain the correlations between the stellar  and supermassive black holes (SMBHs) masses seen within spheroids (e.g.\ di Matteo et al.\  2005), if they operate at the epoch where the bulk of the stars in these galaxies are formed.

Evidence for one AGN-driven feedback mechanism comes from the discovery in the X-ray images of low-redshift galaxy clusters of evacuated bubbles, or cavities,  produced by relativistic particles in radio jets driven by AGN in the central cluster galaxies  (e.g.\ Birzan et al.\ 2004).  Thus powerful radio jets are effective at transfering mechanical energy into the dense intracluster medium (ICM) at low redshift. The redistribution of gas, and thermalization of shock and sound waves may then limit catastrophic cooling in the cluster cores (e.g.\ Allen et al.\ 2006), as predicted for massive galaxies forming at high redshift. While there are  similarities between this jet-driven AGN feedback and that invoked to fix galaxy formation models, the nature of the high-redshift feedback mechanism remains enigmatic: is it mechanical heating of halo gas, or other processes such as winds from merging SMBHs?

In a 130-ks {\it Chandra} observation, Scharf et al.\ (2003, S03) detected X-ray emission across a 100-kpc region around the powerful radio galaxy 4C\,41.17 ($z=3.8$) with a luminosity $\sim 10^{45}$\,erg\,s$^{-1}$.  The spectrum of the emission is non-thermal and matches the power-law index in the radio indicating that this extended X-ray emission arises from inverse Compton (IC) scattering of submillimeter photons from the Cosmic Microwave Background (CMB), or far-infrared photons, by a relativistic electron population associated with current and past activity of the SMBH (S03). 

There is evidence of extended X-ray emission, similar to that seen in 4C\,41.17, around $z\sim2$ radio galaxies (e.g.\ Fabian et al.\ 2003, 2009; Erlund et al.\ 2008; Johnson et al.\ 2007).   These  X-ray structures are  the analogs of the bubble-like cavities seen in low-redshift galaxy clusters, except now the IC emission from the cavities is {\it more} luminous than the thermal cluster emission  (which is cosmologically dimmed).  The intense IC-generated X-ray flux may also help photo-ionize cooling gas around these galaxies (S03).  This would be a new and previously unappreciated feedback mechanism with three key advantages: it preferentially affects 1)  gas on large scales, 2) within the most massive halos, 3) at high redshifts (owing to the rapid evolution of the CMB energy density, $(1+z)^4$).  Such a targeted mechanism, operating only in those halos with SMBHs large enough to power luminous radio jets, is exactly what is required to reduce the gas cooling onto the most massive galaxies, and so ensure that theoretical predictions of their stellar content are in agreement with observations.  The mass-specific nature and redshift-dependency of this feedback mechanism may thus provide a new avenue for theoretical attempts to model the growth of the most massive galaxies seen in the local Universe (Bower et al.\ 2006).

If IC-powered X-ray halos  are going to influence the cooling and star formation history of high-redshift radio galaxies (HzRGs), and thus the formation of giant ellipticals, then they must be prevalent at high redshifts, $z>3$, where the bulk of the star formation in HzRGs occurs (Archibald et al.\ 2001). However, only one $z>3$ HzRG has been observed by {\it Chandra} with sufficient exposure time to unambigiously identify IC emission (S03).  For these reasons, we obtained a 100-ks {\it Chandra} observation of the luminous radio galaxy 4C\,60.07 at $z=3.79$.  This galaxy is undergoing a massive starburst (Stevens et al.\ 2003; Ivison et al.\ 2008, I08) and is surrounded by  an extended Ly-$\alpha$ halo (Reuland et al.\ 2003, 2007).   It thus provides an ideal test of the existence of IC-powered X-ray emission around HzRGs and their influence on the Ly-$\alpha$ halos and gas cooling in their environs.

In our analysis we assume a cosmology with $\Omega_m=0.27$, $\Omega_\Lambda=0.73$ and $H_o=71$\,km\,s$^{-1}$\,Mpc$^{-1}$, giving an angular scale of 7.2\,kpc per arcsec.\ at $z=3.79$.

%
%
\begin{figure*}[tbh]
\centerline{\psfig{file=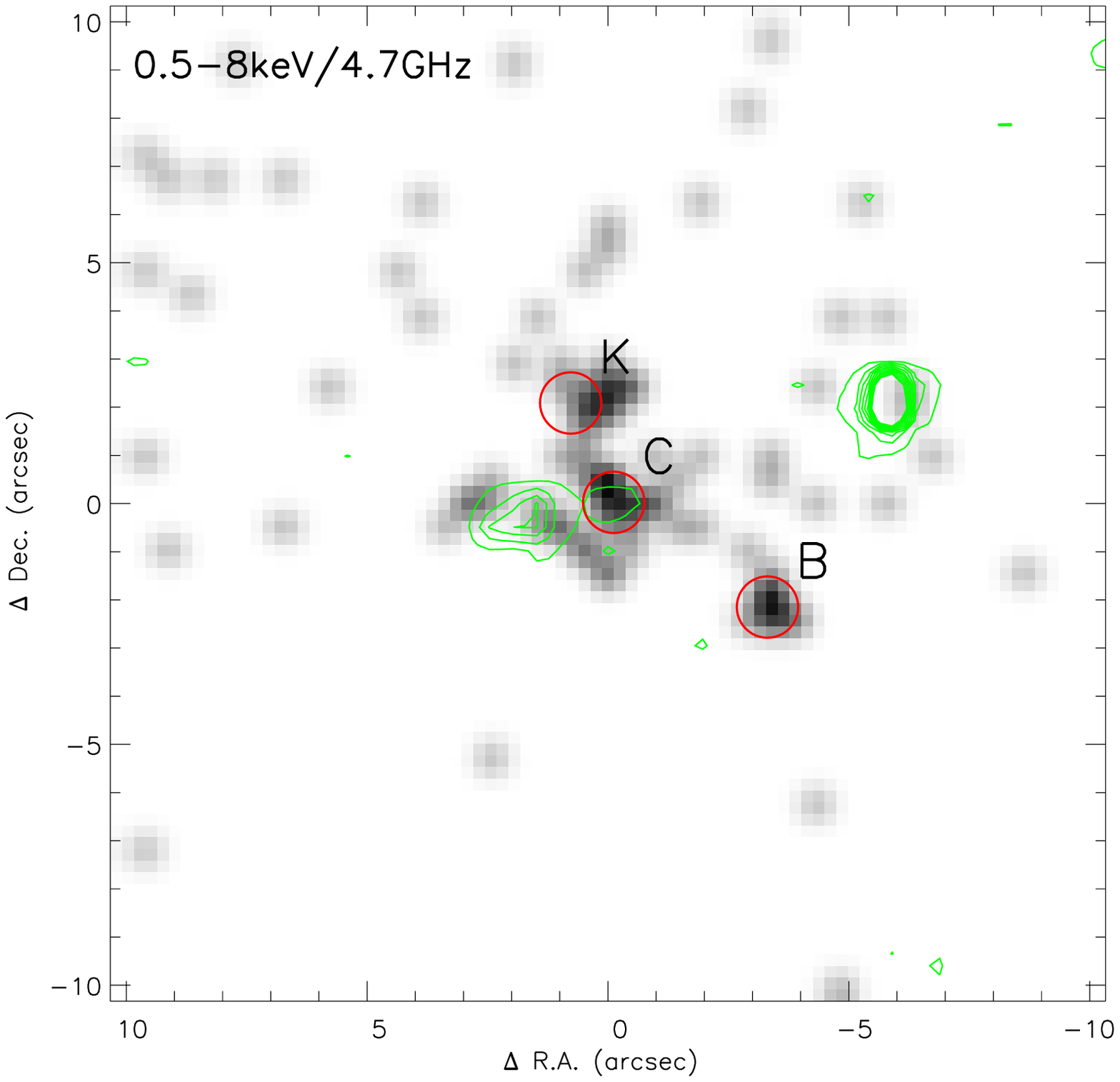,width=2.6in,angle=0}
\hspace*{-0.7cm}
\psfig{file=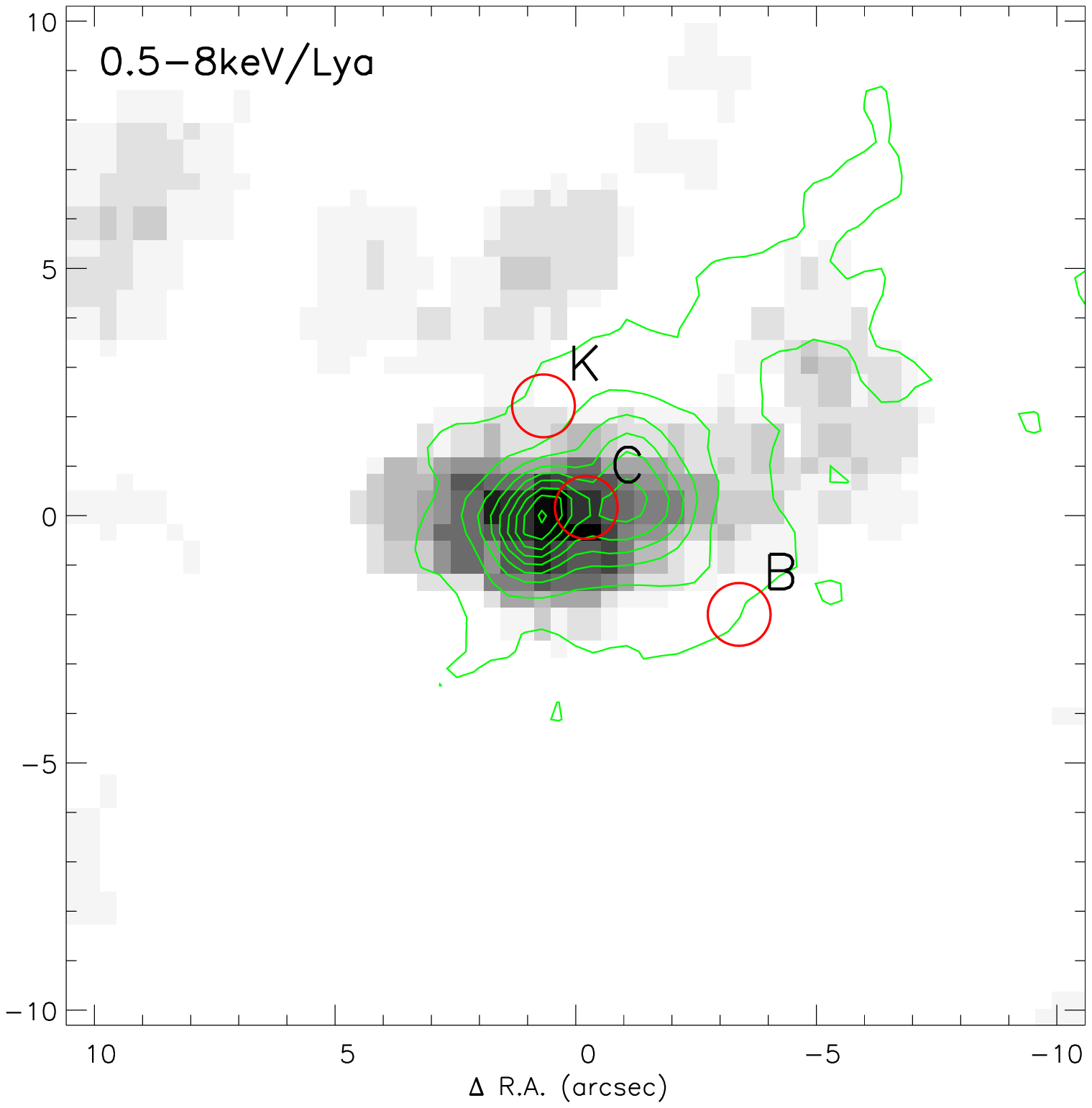,width=2.6in,angle=0}}
\vspace*{-0.7cm}\centerline{
\psfig{file=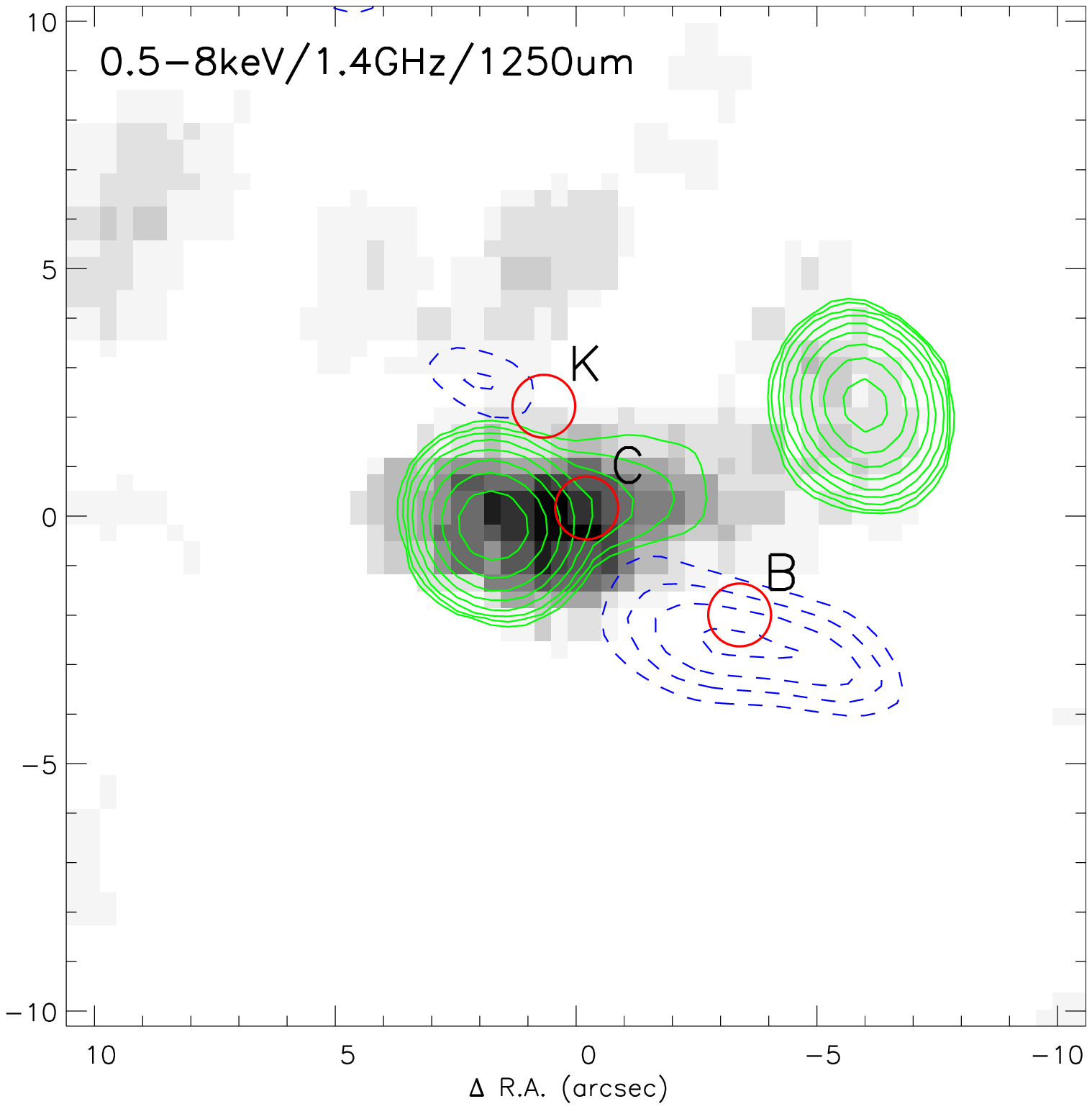,width=2.6in,angle=0}
\hspace*{-0.7cm}
\psfig{file=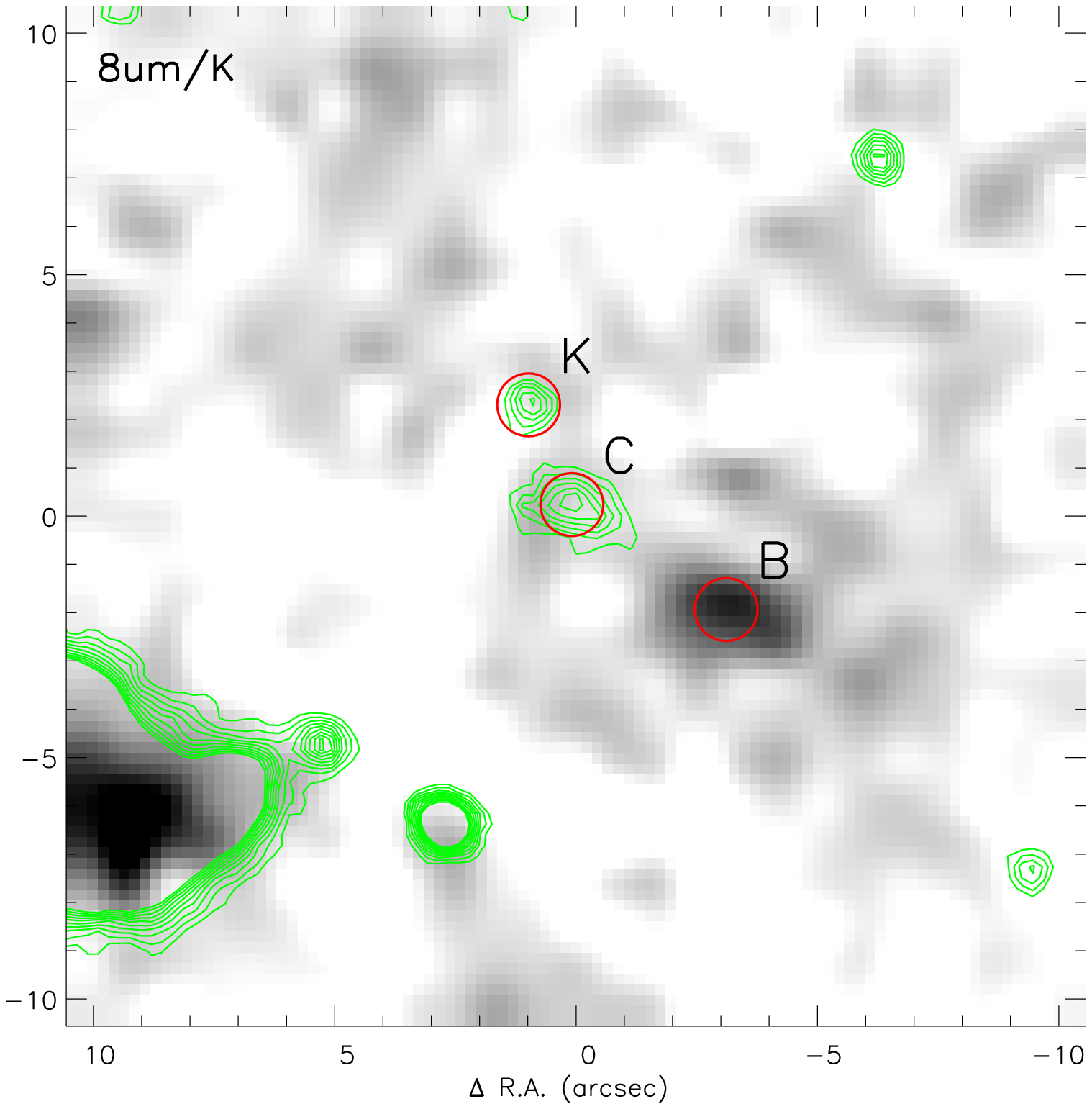,width=2.6in,angle=0}}
\caption{\footnotesize {\it [upper-left]} The {\it Chandra} 0.5--8-keV X-ray image of 4C\,60.07 with the 4.7-GHz VLA map as contours (showing the morphology of the radio jets and core). We mark on the radio core (C) and the two sources B and K  (I08; van Breugel et al.\ 1998).  B is detected at 870\,$\mu$m and is classified as an AGN from its power-law mid-infrared colors (I08).  K is very red in $(I-K)$ and is detected out to 4.5\,$\mu$m, but not at longer wavelengths (I08). Both B and K appear to coincide with X-ray point sources, identifying K as an absorbed AGN and confirming that B is also an AGN. The X-ray image is convolved with a 3.5-kpc FWHM ($0.5''$) Gaussian for display purposes. {\it [upper-right]} The smoothed {\it Chandra} 0.5--8-keV X-ray image after the removal of emission from K, C and B. We plot on this the contours of the Ly-$\alpha$ emission from Reuland et al.\ (2003).  We see that the extended X-ray emission peak coincides with the bright eastern Ly-$\alpha$ emission, while the western extent of the X-ray emission appears to demarcate, but is not coincident with, faint Ly-$\alpha$ emission.  The X-ray image has been convolved with an  18-kpc diameter ($2.5''$) top hat filter.  {\it [lower-left]} The smoothed X-ray image from the upper-right panel, here overlayed with a logarithmic contours from the 1.4-GHz VLA radio map (solid contours) and the 1250-$\mu$m IRAM image (dashed contours, at 2.5, 3, 4, 5-$\sigma$ intervals). {\it [lower-right]} The IRAC 8-$\mu$m image overlayed with Keck $K$-band as contours, illustrating the very different rest-frame $B$- and $H$-band emission of the AGN K and B (see also I08).
}
\end{figure*}

\section{Observations and Reduction}

We obtained a deep {\it Chandra} exposure consisting of  a single $16.9' \times 16.9'$ ACIS-I pointing (Obs-ID 10489)  on 2008 December 10--11, centered close to 4C\,60.07 at 05\,12\,44.48 +60\,29\,58.9 (J2000). The  on-source exposure time was 98.8\,ks. The {\it Chandra} X-ray Center pipeline software (version 7.6.11) was used for the basic data processing.  The reduction and analysis followed Luo et al.\ (2008), including replacing the standard bad-pixel file with one only selecting obvious bad columns and pixels above 1\,keV, which excludes just $\sim 1.5\%$ of the total effective area. We registered the observations by first running {\sc wavdetect} to generate an initial source list and matching these to counterparts in the 3.6-$\mu$m {\it Spitzer} IRAC image (which itself is aligned to the FK5 coordinate frame using 2MASS). This confirmed the absolute astrometry of our image is good to $\sim 0.4''$. Finally we constructed images and exposure maps using the standard {\it ASCA} grade set ({\it ASCA} grades 0, 2, 3, 4, 6) for three  bands:  0.5--8.0\,keV (full band), 0.5--2.0\,keV (soft band), and 2--8\,keV (hard band). The effective exposure time at the position of 4C\,60.07 was 73.0\,ks.  We show the full-band {\it Chandra} image of this region in Fig.~1.   The Galactic absorbing column in this region on degree scales is $2\times10^{21}$\,cm$^{-2}$ (Stark et al.\ 1992). We adopt a minimum uncertainty of 50\% on this column, but caution that this region has structured 100-$\mu$m emission on arcmin.\ scales and so it is possible that the absorption towards 4C\,60.07 is higher than assumed.

\section{Analysis and Results}

As Fig.~1 shows,  extended X-ray emission is visible in our {\it Chandra} image surrounding the radio galaxy out to 5--$8''$ (35--60\,kpc).   There are also at least three point sources within this region.  These correspond to the core of the radio galaxy (C) and two sources previously catalogued at near- and mid-infrared wavelengths: B and K (I08; van Breugel et al.\ 1998).  We note that the X-ray source in K is offset by $\sim 1''$ from the $K$-band counterpart, but this is not highly significant given the X-ray signal to noise and astrometric errors.  We mask these three sources in our analysis of the extended component, replacing them with their local background value.

After removal of the point sources, we see faint extended emission spanning $\sim 12''$ ($\sim 90$\,kpc, Fig.~1) elongated along the same axis as the radio jets and asymmetrically centered on 4C\,60.07.  The profile of the X-ray emission appears to be centrally peaked (even after removal of the central point source; see below) with the extension to the west having more uniform surface brightness. We measure a total of 35 counts in the 0.5--8\,keV image within a $11''\times 4.5''$ ($80\times 32$\,kpc) aperture matched to the emission, after masking the point sources.  This compares to an expected background of just $3.4\pm 1.4$ counts  from   apertures randomly placed across the image.  Correcting for this background we estimate count rates for the extended component of $4.3\pm 0.8 \times 10^{-4}$\,cts\,s$^{-1}$  (0.5--8\,keV), $1.7\pm 0.5 \times 10^{-4}$\,cts\,s$^{-1}$  (0.5--2\,keV) and  $2.6\pm 0.7 \times 10^{-4}$\,cts\,s$^{-1}$ (2--8\,keV).  These translate into a 0.5--8\,keV flux of $8.2\pm 1.6\times 10^{-15}$\,erg\,s$^{-1}$\,cm$^{-2}$  after correcting for Galactic absorption and assuming a power-law spectrum corresponding to our observed 2--8/0.5--2\,keV band-ratio of  $1.6^{+1.3}_{-0.7}$, equivalent to a photon index ($\Gamma=1-\alpha$, where $S_\nu \propto \nu^\alpha$)  of $\Gamma_{\rm eff}=0.8_{-0.7}^{+0.6}$ including our nominal uncertainty in the Galactic absorption.   The extended component is too faint to yield a detailed X-ray spectrum (c.f.\ S03). 

Assuming the extended emission has an intrinsic photon index typical of obscured AGN, $\Gamma_{int}=1.9$, the observed band ratio implies an intrinsic absorption of  $5\times 10^{23}$\,cm$^{-2}$.  This absorption would need to extend over $\sim 2500$\,kpc$^2$, requiring a mass in the absorber of $10^{13}$\,M$_\odot$.  We do not believe it is feasible to have such an extended and massive absorber.  Instead we assume that the observed spectral index is intrinsic to the emission and derive an X-ray luminosity in the observed 0.5--8\,keV band (rest-frame 2.4--38.4\,keV) of $L_{\rm X}=12\pm 2\times 10^{44}$\,erg\,s$^{-1}$ or $L_{\rm X}=1.8\pm 0.2\times 10^{44}$\,erg\,s$^{-1}$ in the rest-frame 0.5--8\,keV band.

Our target lies near the aim-point of the {\it Chandra} observations and so the telescope optics deliver impressive image quality, $\sim 1''$ FWHM.  We exploit this to derive fluxes for the three point sources around 4C\,60.07 using small, $2''\times2''$, apertures.  We measure total counts in the 0.5--8\,keV (0.5--2\,keV, 2--8\,keV) band of 13 (4, 9), 6 (3, 3) and 9 (6, 3) for K, C and B respectively.  We convert these into count rates using the exposure maps for each band and then subtract the local background estimated from smoothing the point-source subtracted map.  The background-corrected 0.5--8\,keV count rates are $1.5\pm 0.4\times 10^{-4}$, $0.8\pm 0.4\times 10^{-4}$  and $1.5\pm 0.5\times 10^{-4}$\,cts\,s$^{-1}$   for  K, C and B respectively, where the uncertainties are simply Poisson estimates.  These translate into 2--8/0.5--2\,keV band-ratios for K, C and B of:  $2.1^{+4.0}_{-1.2}$,  $1.0^{+2.9}_{-0.7}$ and  $0.5^{+0.8}_{-0.3}$, and the corresponding photon indices are $\Gamma_{\rm eff}= 0.5\pm 0.6$, $1.3\pm 1.1$ and $2.0\pm 1.0$  (corrected for Galactic absorption).  Assuming that K lies at $z\sim 3.8$ (the same redshift as C and B, I08), we find that the X-ray color of B is consistent with it being unabsorbed, while C and K likely have moderate intrinsic absorption of $N({\rm H}{\sc i})\gs 10^{23}$\,cm$^{-2}$  (adopting an intrinsic photon index of $\Gamma_{int}=1.9$).   The corresponding X-ray luminosities are  $2.1\times 10^{44}$, $0.8\times 10^{44}$ and $2.1\times 10^{44}$\,erg\,s$^{-1}$ in the observed 0.5--8\,keV band, or  $1.8\times 10^{44}$, $1.0\times 10^{44}$ and $1.8\times 10^{44}$\,erg\,s$^{-1}$ at  rest-frame 0.5--8\,keV.

In our analysis we also make use of archival multiwavelength observations of 4C\,60.07.  To derive a far-infrared luminosity we have rereduced the 450-$\mu$m SCUBA map from I08 and derived a flux of $28.3\pm 7.5$\,mJy for the region around 4C\,60.07.  Fitting a  modified black body with $\beta=1.5$ to this and the 850, 890 and 1250$\mu$m  fluxes from I08 and Papadopoulos et al.\ (2000) yields $T_d= 36\pm 4$\,K and $L_{FIR}= 1.0^{+1.5}_{-0.3}\times 10^{13}L_\odot$. Assuming instead the dust temperature measured for the well-sampled dust spectrum of 4C\,41.17, we derive $L_{FIR}\sim 2\times 10^{13}L_\odot$.  We also convert the Ly-$\alpha$ luminosity for 4C\,60.07 from Reuland et al.\ (2003) to our cosmology and derive  $L_{Ly\alpha}=1.2\times 10^{45}$\,erg\,s$^{-1}$.

For comparison, we have also converted the properties of the extended X-ray emission  around  4C\,41.17 to the X-ray bands used here.  The equivalent values are  $\Gamma_{\rm eff}= 1.57\pm 0.26$, a 0.5--8\,keV flux of $8.5\pm 0.9 \times 10^{-15}$\,erg\,s$^{-1}$\,cm$^{-2}$, an X-ray luminosity in the observed 0.5--8\,keV (rest-frame 2.4--38.4\,keV) of $L_{\rm X}=12\pm 3\times 10^{44}$\,erg\,s$^{-1}$ and $L_{\rm X}=5.9\pm 1.4\times 10^{44}$\,erg\,s$^{-1}$ in the rest-frame 0.5--8\,keV band.  Similarly we derive $T_d=45\pm 1$\,K and  $L_{FIR}=1.6\pm 0.4 \times 10^{13}L_\odot$ from fitting a modified black body with $\beta=1.5$ to the extensive sub-millimeter photometry for 4C\,41.17 tabulated by Greve et al.\ (2007).  The Ly-$\alpha$ luminosity of 4C\,41.17 is $L_{Ly\alpha}=1.2\times 10^{45}$\,erg\,s$^{-1}$ from Reuland et al.\ (2003) and converted to our cosmology, identical to that of 4C\,60.07.

\section{Discussion and Conclusions}

We have discovered extended X-ray emission around the HzRG 4C\,60.07 at $z=3.79$.  This is only the second example of extended X-ray emission known at $z>3$ (after 4C\,41.17, S03).  However, instead of rarity, we believe that this simply reflects the lack of sufficiently deep {\it Chandra} observations of HzRGs at these redshifts and that such extended X-ray emission is actually common around massive radio galaxies at $z>3$.

Comparing the various emission components around 4C\,60.07 and 4C\,41.17 (S03), both  at $z=3.8$, we see that they have identical  rest-frame 1.7-GHz luminosities (observed 365-MHz): $L_{\rm 1.7GHz}=1.4\times 10^{44}$\,erg\,s$^{-1}$.  The extended Ly-$\alpha$ halos also have identical luminosities: $L_{Ly\alpha}=1.2\times 10^{45}$\,erg\,s$^{-1}$ and the X-ray luminosities of their similar-sized halos ($\sim 90$--100\,kpc) are also identical (in the observed 0.5--8\,keV band): $L_{\rm X}=12\pm 2\times 10^{44}$\,erg\,s$^{-1}$ and $L_{\rm X}=12\pm 3\times 10^{44}$\,erg\,s$^{-1}$ respectively.  The far-infrared luminosities of these systems are also comparable: $L_{FIR}=1.0^{+1.5}_{-0.3} \times 10^{13}L_\odot$ for 4C\,60.07 (likely towards the upper-end of this range) and $L_{FIR}=1.6\pm 0.4 \times 10^{13}L_\odot$ for 4C\,41.17 and both show evidence for either multiple far-infrared sources or spatially extended emission.  The only apparent difference between the two HzRGs is the photon index of the X-ray emission: $\Gamma_{\rm eff}\sim 0.8_{-0.7}^{+0.6}$ for 4C\,60.07, compared to $\Gamma_{\rm eff}= 1.57\pm 0.26$ for 4C\,41.17, although at $\sim 1.3$-$\sigma$ this is not statistically significant, especially given the possibility of structured Galactic absorption towards 4C\,60.07 which would suppress $\Gamma_{\rm eff}$. Therefore we take the close agreement between the X-ray, GHz-radio and Ly-$\alpha$ emission in these two HzRGs as a strong indication that these three components are linked by a single physical process.

The scale of the X-ray emission we find in 4C\,60.07, across a $\sim 90$\,kpc region with a luminosity-weighted radius of the X-ray emission of $\sim 20$\,kpc, is comparable to the size of both the radio and Ly-$\alpha$ emission (Fig.~1; Reuland et al.\ 2003).  However, the detailed morphology of the extended X-ray component appears to be better matched to the radio emission than the Ly-$\alpha$ (Fig.~1).  This hints at a causal relationship between the radio and X-ray emission, which, given the similarity to the equivalent components in 4C\,41.17 (S03), in turn suggests IC scattering as the most likely mechanism.

The radio spectral index of the jets in 4C\,60.07 at $>1$-GHz is $\alpha \sim -1.6$ and becomes shallower at lower frequencies, $\alpha \sim -0.9$ at $\sim 100$\,MHz in the observed frame (Chambers et al.\ 1996; I08).  The CMB, and potentially the luminous dusty starbursts in and around 4C\,60.07 (I08), provide a source of submillimeter/far-infrared photons which scatter off the relativistic electron population in these jets.  Using the observed flux and size of the 1.4-GHz radio emission (Fig.~1) and a path length through the jet equal to its transverse size, $\sim 40$\,kpc, we derive a minimum-energy magnetic field in the radio source (equivalent to the equipartition field) of $\sim 50\mu$G (see Eq.~3, Miley 1980). Assuming this field, electrons with $\gamma\sim 1500$ will scatter $z=3.8$ CMB photons up to X-ray energies of 2\,keV (rest-frame $\sim 10$\,keV) and are also responsible for synchrotron radio emission at $\sim 0.5$\,GHz (rest-frame $\sim 2.5$\,GHz).  The similarity between the large scale X-ray emission and the (relatively low-resolution) 1.4-GHz map in Fig.~1 is consistent with these two components arising from the same electron population.  

As noted earlier, the far-infrared luminosity arising in 4C\,60.07 and its companions is similar to 4C\,41.17 and so, as shown by S03, we expect the far-infrared and CMB photon densities to be comparable within the central few 10's kpc of the source (dependent upon the distribution of the far-infrared emission and the jet plasma).  The energy density in the CMB at $z=3.79$ is $\rho_{CMB}=2\times 10^{-10}$\,erg\,cm$^{-3}$;  assuming a homogeneous distribution for the far-infrared emission of 4C\,60.07 the far-infrared from the starburst will exceed the CMB photon density within a radius of $\sim 50$\,kpc. Moreover, 4C\,60.07's dust emission has a characteristic temperature of $\sim 36$\,K, only slightly hotter than the CMB at this redshift, $T\sim 13$\,K.  X-ray emission at 2\,keV thus results from scattering of these far-infrared photons from electrons with $\gamma\sim 1000$, whose synchrotron emission peaks at $\sim 200$\,MHz (or $\sim 40$\,MHz in the observed frame). This electron population is more abundant than that responsible for the IC-scattering the CMB and thus potentially dominates the IC emission in the central regions.  As with 4C\,41.17, there are hints of an $r^{-2}$-like component in the X-ray morphology, with a flatter emission profile on larger scales, which is consistent with both a compact far-infrared source and the uniform CMB illuminating the halo. 

Our X-ray data also identify three luminous AGN within a small region around the HzRG, this is a very rare configuration (Smail et al.\ 2003).  Two of these AGN definitely lie at $z=3.79$:  B and the radio galaxy core (C), from their CO line emission (Papadopoulos et al.\ 2000; I08).  The third AGN, K, lies within 15\,kpc of the radio galaxy core and has a likelihood of being a chance projection of just 0.25\% based on its X-ray flux.  Moreover, the near- and mid-infrared spectral energy distribution of K is similar to the HzRG (Fig.~1 and I08) strengthening the likelihood that it is at $z=3.79$.  Fig.~1 shows that B is a strong millimeter source, while potential millimeter emission may also be associated with K, suggesting that dusty starbursts are occuring in both these companions. The X-ray luminosities of these AGN suggest SMBH masses of $\gs 10^7$\,M$_\odot$ if their accretion is Eddington-limited, indicating that these are both reasonably massive ($\sim 10^{10}$\,M$_\odot$) and relatively evolved host galaxies. The two companion AGN are within 30-kpc in projection from the HzRG.  At 300\,km\,s$^{-1}$ this distance would take just $\sim 100$\,Myrs to cover, similar to the likely age of the radio source and the age of the starburst. This suggests that both types of activity were dynamically triggered by the interactions we are seeing.
\smallskip

The main conclusions of this work are:

1. We have identified X-ray emission extending over a $\sim 90$-kpc region around the $z=3.79$ luminous radio galaxy, 4C\,60.07.  This is only the second known example of extended X-ray emission around a HzRG at $z>3$.   

2. The very close similarity of the extended X-ray emission around 4C\,60.07 to that seen around 4C\,41.17 (also at $z=3.8$) by Scharf et al.\ (2003), argues that it arises from IC scattering of CMB and far-infrared photons from relativistic electrons  in the radio jets  (as in 4C\,41.17).  This process may be a significant source of heating for the gaseous halos around these massive galaxies at high redshifts, and should be incorporated in theoretical models of their formation.

3. Our deep {\it Chandra} X-ray observations also identify three luminous AGN in the halo of 4C\,60.07.  One of these corresponds to the core of the radio galaxy and a second is a previously mid-infrared-identified AGN now confirmed from its X-ray emission.  The final X-ray identified AGN appears to be coincident with a near-infrared counterpart (van Breugel et al.\ 1998).  The presence of three far-infrared-luminous SMBHs in a $\sim 30$\,kpc region is strong evidence for the triggering of radio and starburst activity in 4C\,60.07 through the merger of multiple massive galaxies.

4. The discovery of both an IC-powered X-ray halo around this high redshift galaxy and evidence for multiple merging SMBHs within its halo demonstrate the wealth of information which high-resolution X-ray observations can yield on the physical processes which drive the formation of massive galaxies at high redshifts.

\acknowledgments
We thank the referee for useful comments and Michiel Reuland, Alastair Edge and Jonathan Gelbord for help.
I.R.S., B.D.L., R.J.I., R.G.B., J.E.G.\ \& K.E.K.C.\ acknowledge support from  STFC. 
D.M.A.\ acknowledges support from the Royal Society.
C.A.S.\ acknowledges support from the NASA/{\it Chandra} grant SAO G09--0152X.

\end{document}